\documentclass[aps,prl,twocolumn]{revtex4}
\usepackage{amssymb}

\usepackage{graphicx}

\begin{document}

\title{Irreversible dynamics of the phase boundary in U(Ru$_{0.96}$Rh$_{0.04}$)$_2$Si$_2$
and implications for ordering}
\author{A.~V.~Silhanek,$^1$ M.~Jaime,$^1$ N.~Harrison,$^1$ V.~Fanelli,$^2$
C.~D.~Batista,$^1$ H.~Amitsuka,$^3$, S.~Nakatsuji,$^4$ L. Balicas,$^4$ K. H.
Kim,$^1$ Z. Fisk,$^4$ J.~L.~Sarrao,$^5$ L. Civale,$^5$ and J. A. Mydosh$^{6}$}
\affiliation{$^1$National High Magnetic Field Laboratory, Los Alamos National Laboratory,
MS E536, Los Alamos, NM 87545, USA\\
$^2$Dep. of Phys. and Astron., University of California, Irvine, CA
92697-4575\\
$^3$Graduate School of Science, Hokkaido University, N10W8 Sapporo
060-0810,Japan\\
$^4$National High Magnetic Field Laboratory, Florida State University,
Tallahassee, Florida 32306\\
$^5$MST Division, Los Alamos National Laboratory, Los Alamos, New Mexico
87545\\
$^6$Max-Planck-Institut f\"{u}r Chemische Physik fester Stoffe, 01187
Dresden, Germany}
\date{\today}

\begin{abstract}
We report measurements and analysis of the specific heat and magnetocaloric
effect at the phase boundary into the single magnetic field-induced phase (phase II) of
U(Ru$_{0.96}$Rh$_{0.04}$)$_2$Si$_2$, which yield striking
similarities to the valence transition of Yb$_{1-x}$Y$_x$InCu$_4$. To explain these similarities, we propose a bootstrap mechanism by which a structural distortion causes an electric quadrupolar order parameter within phase II to become coupled to the $5f$-electron hybridization, giving rise to a valence change at the transition.
\end{abstract}

\pacs{PACS numbers: ..............................}
\maketitle

The broken symmetry order parameter responsible for the large specific heat anomaly at $T_{\rm o}\sim$~17~K in URu$_{2}$Si$_{2}$ continues to be of interest owing to its elusive `hidden' nature~\cite{matsuda1,behnia1}. While there has been no clear consensus on the appropriate theoretical description of the `hidden order' (HO) phase~\cite{libero1,barzykin1,santini1,ohkawa1,chandra1,virosztek1,fazekas1,mineev1,varma1,tripathi1}, the key properties of the Fermi liquid state, upon which the HO parameter manifests itself, can be understood. Comprehensive de~Haas-van~Alphen (dHvA) measurements~\cite{ohkuni1} reveal the quasiparticles to have heavy effective masses and Ising spin degrees of freedom~\cite{silhanek2}, indicating that nearly-localized $5f$-electron degrees of freedom contribute to the Fermi liquid. Within an Anderson lattice scheme, hybridization causes the itinerant quasiparticles to acquire the spin and orbital degrees of freedom of a lowest lying crystal electric field $5f^2$-multiplet~\cite{park1}.  In this case, it is a  $\Gamma_5$ non-Kramers doublet~\cite{ohkawa1}, making the quasiparticles in URu$_2$Si$_2$ unique among heavy fermion materials, possessing electric quadrupole as well as Ising spin degrees of freedom~\cite{silhanek2}. 

On the basis of the dHvA experiments, one can further assert that the HO phase I  (that exists at magnetic fields $\mu_0H\lesssim$~35~T and temperatures $T\lesssim$~17~K~\cite{jaime1}) must be explained within the context of an itinerant $\Gamma_5$ quasiparticle model. This is required to account for the survival of $\Gamma_5$ quasiparticles deep within the HO phase~\cite{ohkuni1,silhanek2,nakashima1}. Itinerancy of the $f$-electrons could also be an important factor in making an itinerant electric-quadrupolar order parameter difficult to detect, as proposed for some of the more exotic itinerant models~\cite{chandra1,virosztek1,varma1,tripathi1}. Given that the HO parameter has defeated attempts at a direct observation, the existence of `$\Gamma_5$ quasiparticles' necessitates an alternative question:  can the electric quadrupolar degrees of freedom order within an itinerant $5f$-electron model, and, if so, how might such ordering differ from the established local moment quadrupolar systems such as UPd$_3$~\cite{mcmorrow1}? 
Until such questions are addressed by a micriscopic theory, an alternative approach to exploring the question of electric quadrupolar order is to tip the balance of the interactions in favor of local moment quadrupolar order of the type seen in UPd$_3$~\cite{ohkawa1,mcmorrow1}. In URu$_2$Si$_2$ this might be achieved in two ways. The first is by Rh-doping, which shifts the spectral weight of the $5f$-electrons away from the Fermi energy~\cite{sandratskii1}, weakening the extent to which they hybridize. Rh-doping also inhibits ${\bf q}$-dependent itinerant mechanisms by smearing the states at the Fermi surface. The second is by applying strong magnetic fields, which enhance the effect of local correlations in the vicinity of the metamagnetic transition, causing the quasiparticle bandwidth to collapse~\cite{silhanek1}. 
The large magnetic susceptibility associated with metamagnetism also strongly favors XY order~\cite{jaime2} (in which electric quadrupole `pseudospins' lie orthogonal to the tetragonal $c$-axis) over Ising antiferromagnetic order~\cite{ohkawa1}. 

\begin{figure}[tbh]
\centering \includegraphics*[scale=0.4]{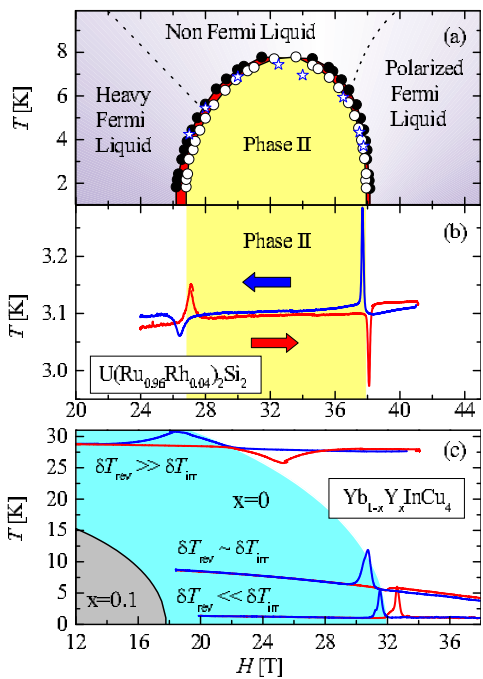}
\caption{(a) Phase diagram of U(Ru$_{0.96}$Rh$_{0.04}$)$_{2}$Si$_{2}$
determined by MCE (circles) and $C_p(T)$ (star symbols). Solid (open)
circles indicate the phase boundary exiting (entering) the ordered phase II. The dashed lines represent the approximate Fermi energy scale, as modified by correlations due to metamagnetism~\cite{silhanek1}. (b) MCE of U(Ru$_{0.96}$Rh$_{0.04}$)$_{2}$Si$_{2}$, with arrows indicating the direction of $H$ sweep.
(c) MCE of YbInCu$_{4}$, with the changes $\delta T$ in $T$ rescaled $\times$2.}
\label{phasediagram}
\end{figure}

In this paper, we propose that the interplay between an electric quadrupolar order parameter and the extent $V_{fc}$ to which $5f$-electrons hybridize causes the transition into phase II in 
U(Ru$_{0.96}$Rh$_{0.04}$)$_{2}$Si$_{2}$~\cite{kim1} (shown in Fig.~\ref{phasediagram}) to acquire thermodynamic similarities to the valence transition of YbInCu$_4$~\cite{dallera1,mushnikov1}. These similarities are evident both in magnetocaloric effect (MCE) and specific heat $C_p(T)$ data. When a more conventional (i.e. non-itinerant) type of electric quadrupolar order~\cite{ohkawa1,mcmorrow1} occurs, the associated structural distortion can alter $V_{fc}$, which determines both the effective `Kondo temperature' and the valence state of the 
system~\cite{dallera1,mushnikov1,drymiotis1}.

%U(Ru$_{0.96}$Rh$_{0.04}$)$_{2}$Si$_{2}$ single crystals are grown using the
%Czochralski method as described elsewhere~\cite{yokoyama1}, while Yb$_{1-x}$Y$_{x}$InCu$_{4}$ 
%are prepared by flux growth with $x=$~0 and $x=$~0.1~\cite{dallera1,mushnikov1}.  All measurements are
%performed at the National High Magnetic Field Laboratory (NHMFL).
%
The MCE is a convenient tool for studying phase boundaries in
a magnetic field~\cite{jaime1}. Here, the sample temperature $T$ is
recorded while the magnetic field $H$ is swept rapidly under quasi-adiabatic conditions.
When an order-disorder transition is crossed, an abrupt change in $T$
reflects the fact that entropy cannot change. A
typical MCE measurement for U(Ru$_{0.96}$Rh$_{0.04}$)$_{2}$Si$_{2}$ is shown
in Fig. \ref{phasediagram}b. The $T(H)$ curve obtained during the $H$ up-sweep 
(red line) shows a sudden increase when $\mu _{0}H\sim $~27~T, indicating that the system
enters an ordered phase: i.e. $T$ must increase in order to conserve entropy. 
The system then relaxes to equilibrium with the bath until the next phase
boundary, exiting phase II, is encountered at $\mu _{0}H\approx $~38~T. Now $T$ 
swings the opposite direction as the ordered phase is abandoned.
The observed MCE anomalies are hysteretic, as evident from their dependence on the direction of $H$ sweep, which is a direct consequence of the transition being of first order. Another systematic feature of the data is that the change in $T$ is larger 
at the high $H$ phase boundary, being consistent with a larger jump in the magnetization~\cite{kim1}. Furthermore, the net magnitude of the swing in $T$ is
larger when entering phase II than exiting it. A similar asymmetry is observed in 
Yb$_{1-x}$Y$_x$InCu$_4$ (with $x=$~0), see Fig. \ref{phasediagram}c.

%% figure phase diagram %%

\begin{figure}[tbh]
\centering \includegraphics*[scale=0.4]{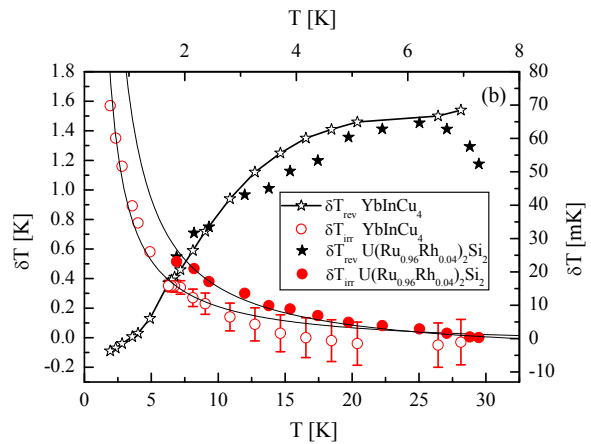}
\caption{Irreversible $\protect\delta T_{\mathrm{irr}}$ (circles) and
reversible $\protect\delta T_{\mathrm{rev}}$ (stars) components of the MCE
as a function of $T$ for U(Ru$_{0.96}$Rh$_{0.04}$)$_{2}$Si$_{2}$ (solid
symbols) and YbInCu$_{4}$ (open symbols). The data corresponding to 
U(Ru$_{0.96}$Rh$_{0.04}$)$_{2}$Si$_{2}$ (YbInCu$_{4}$) are linked to the right
(left) and upper (lower) axes. The thin lines are fits of $E_{\rm p}(T)$ to the $\protect\delta T_{\mathrm{irr}}$ data while the thick
lines are merely guides to the eye.}
\label{analysis}
\end{figure}

The total swing $\delta T$ in $T$ in the MCE curve (for a given
direction of $H$ sweep) is the sum of reversible $\delta T_{\mathrm{rev
}}$ and irreversible $\delta T_{\mathrm{irr}}$ contributions. 
$\delta T_{\mathrm{rev}}$ always changes sign when the direction of sweep of 
the magnetic field is changed, while $\delta T_{\mathrm{irr}}$ does not, since is entirely due to
irreversible (or dissipative) processes. The latter can be associated with the pinning of domain boundaries 
to the crystalline lattice~\cite{blab0}, causing the system to become
metastable with its actual state depending on its history. Pinning
forces become especially relevant if the order parameter involves charge
degrees of freedom, as is known to be the case for valence transitions and electric quadrupolar phases~\cite{mushnikov1, ohkawa1}.
Figure~\ref{analysis} shows the extracted $\delta T_{\mathrm{rev}}$ and 
$\delta T_{\mathrm{irr}}$ for both U(Ru$_{0.96}$Rh$_{0.04}$)$_2$Si$_2$ and 
Yb$_{1-x}$Y$_x$InCu$_4$ (with $x=$~0). In the case of 
U(Ru$_{0.96}$Rh$_{0.04}$)$_2$Si$_2$, data is shown only for the upper critical field and the axes
have been rescaled to
compare the two systems. At higher $T$, $\delta T_{\mathrm{rev}}$ dominates the MCE
in both materials, but the entropy vanishes as $T\rightarrow$~0, $\delta T_{\mathrm{rev}}$ must also vanish, causing $\delta T_{\rm irr}$ to dominate in that limit. The latter grows
rapidly as $T\rightarrow$~0: this is especially clear in the case of
YbInCu$_4$ for which a greater range in $T$ can be accessed owing to its high transition temperature. 

%% figure Magneto Caloric %%

\begin{figure}[tbh]
\centering \includegraphics*[scale=0.4]{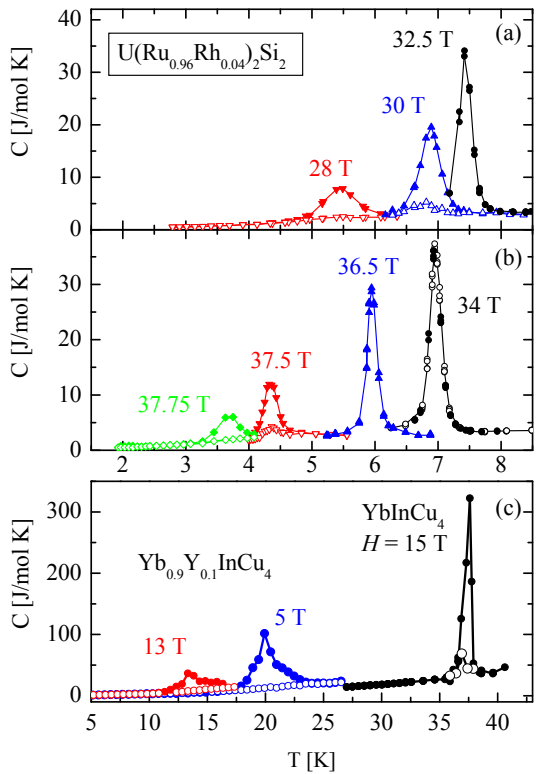}
\caption{$C_p(T)$ at different values of
$H$ measured on (a) 
U(Ru$_{0.96}$Rh$_{0.04}$)$_{2}$Si$_{2}$ for $\protect\mu _{0}H<$~34~T, (b) 
U(Ru$_{0.96}$Rh$_{0.04}$)$_{2}$Si$_{2}$ for $\protect\mu _{0}H\geq $~34~T, and 
(c) Yb$_{1-x}$Y$_{x}$InCu$_{4}$(with $x=$~0.1) using a superconducting magnet. Solid symbols represent 
$C_p$ measured with the first heat pulse on warming-up while open symbols
represent the average of subsequent heat pulses. }
\label{specificheat}
\end{figure}

The similarities between U(Ru$_{0.96}$Rh$_{0.04}$)$_2$Si$_2$ and 
Yb$_{1-x}$Y$_x$InCu$_4$ also extend to measurements of the $C_p(T)$. 
Figure~\ref{specificheat} shows $C_p(T)$ for U(Ru$_{0.96}$Rh$_{0.04}$)$_2$Si$_2$
and Yb$_{1-x}$Y$_x$InCu$_4$ (with $x=$~0 and 0.1) measured at constant $H$ at many different temperatures using the thermal 
relaxation time method~\cite{jaime1} (both during a warm up and cooling down of the sample 
using a small $\sim$~1-3\% $T$ increment). 
The first $C_p(T)$ point measured at each $T$ during the warm up yields a larger value 
(solid symbols) than subsequent points (open symbols), which is consistent with the irreversibilities observed using the MCE. However, neither the
first $C_p(T)$ point (as explained above) nor the subsequent points can be used to 
extract the precise entropy change at the transition. The former includes the energy absorbed by irreversible processes (depinning domain boundaries etc.) in addition to the equilibrium
$C_p(T)$. Once the sample has settled into a new metastable state at each 
$T$, subsequent measurements have a much reduced effect on its state, leading
to a smaller estimate for $C_p(T)$. Type-II superconductors are well known to give rise to a similar irreversible behaviour~\cite{whatsthis}, with the pinned current profile relaxing considerably after the initial thermal cycle.

%% figure Phase Diagram %%

It is, nevertheless, worth emphasizing the primary difference between 
U(Ru$_{0.96}$Rh$_{0.04}$)$_2$Si$_2$ and Yb$_{1-x}$Y$_x$InCu$_4$. The valence instability in Yb$_{1-x}$Y$_x$InCu$_4$ is the consequence of a situation whereby the total free energy of the solid develops multiple minima as a function of the volume along with $V_{fc}$~\cite{dallera1,mushnikov1}. One can consider the mean value of the hybridization operator as the effective order parameter, although the $f$-electron density is the most natural order parameter to describe a valence transition. Such an order parameter is non symmetry-breaking, making it directly analogous to boiling point of a liquid.
The fact that phase II in U(Ru,Rh)$_2$Si$_2$ occurs only under finite magnetic fields in the vicinity of a metamagnetic crossover~\cite{silhanek1}, around which the correlations are strongly enhanced, indicates that this phase has a different origin to the valence transition in Yb$_{1-x}$Y$_x$InCu$_4$ or Ce$_{0.8}$Th$_{0.1}$La$_{0.1}$~\cite{dallera1,mushnikov1,drymiotis1}.

Phase II in U(Ru$_{0.96}$Rh$_{0.04}$)$_2$Si$_2$ can condense at temperatures that exceed the characteristic Fermi liquid temperature (extrapolated from outside the ordered phase)~\cite{silhanek1}. This together with the roughly symmetric shape of the phase boundary around the metamagnetic transition is strongly suggestive of local moment ordering. A broken symmetry order parameter involving a structural distortion of the lattice, so as to change $V_{fc}$ within phase II, would provide a very effective bootstrap mechanism for both ensuring its stability and causing the transition to become first order like that in YInCu$_4$. The large changes in sound velocity observed by Suslov {\it et al.}~\cite{suslov1} in pure URu$_2$Si$_2$ indicate a pronounced magnetoelastic coupling consistent with a lattice distortion. Furthermore, the large redistribution of entropy involved in the formation of phase II indicates that the $5f$-electrons are involved~\cite{silhanek1}, which strongly favors a lattice distortion caused by a electric quadrupolar order parameter as opposed to a charge-density wave. 

The possibility of a broken symmetry order parameter being coupled to $V_{fc}$ may have wider appeal than U(Ru$_{0.96}$Rh$_{0.04}$)$_2$Si$_2$. For example, an equivalent coupling in YbRh$_2$Si$_2$ would provide a rather natural explanation as for why the critical field of its unidentified low $T$ ordered phase acquires the physical characteristics of a `valence fluctuator' quantum critical point~\cite{norman1}.

\begin{figure}[tbh]
\centering \includegraphics*[scale=0.45]{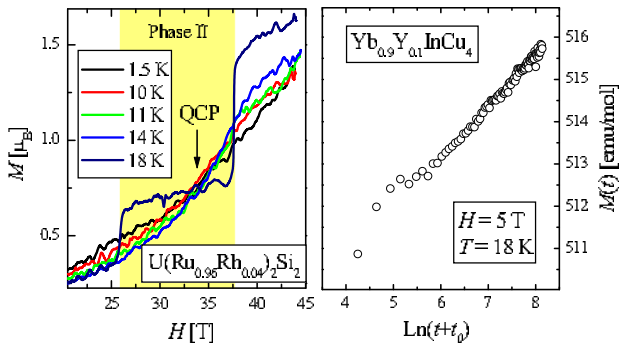}
\caption{(a) magnetization as a function of field at several
temperatures for the U(Ru$_{0.96}$Rh$_{0.04}$)$_{2}$Si$_{2}$ sample. (b) Relaxation of the magnetization for the 
Yb$_{0.9}$Y$_{0.1}$InCu$_{4}$ sample at $\protect\mu _{0}H=$~5~T and $T=$~18~K plotted as an
stretched exponential.}
\label{metastable}
\end{figure}

Having established that the $f$-electron valence plays an equally important role in dominating the thermodynamics of both U(Ru$_{0.96}$Rh$_{0.04}$)$_2$Si$_2$ and Yb$_{1-x}$Y$_x$InCu$_4$, further analysis of the irreversible processes are required in order to understand Fig.~\ref{analysis}. The increase in $\delta T_{\rm irr}$ as $T\rightarrow$~0 is consistent with the loss of thermal fluctuations, which enable the domain boundaries to overcome pinning forces and undergo creep at finite $T$. Type II superconductors provide a good analogue for understanding creep~\cite{blatter1}, with the supercurrents sustained by pinned vortices being replaced in the present valence systems by the magnetic currents associated with the difference in magnetization $\Delta M$ between domains. To model the present experimental data, we introduce a phenomenological model for the energy 
$E_{\mathrm{p}}\propto\exp(U_0/k_{\mathrm{B}}T)-\exp(U_0/k_{\mathrm{B}}T_{\rm o})$
stored due to pinning. Here, $U_0$ is the typical energy of a pinning site~\cite{blatter1}, while $T_{\rm o}$ is the characteristic ordering temperature introduced to constrain the model so that $E_{\rm p}$ vanishes when $\Delta M$ vanishes. In both systems, the transition temperature $T_{\rm o}$ is optimal (maximum) when $\Delta M=$~0, occurring at $\approx$~34~T in U(Ru$_{0.96}$Rh$_{0.04}$)$_2$Si$_2$ (see Fig.~\ref{metastable}a) and at $H=$~0 in
Yb$_{1-x}$Y$_x$InCu$_4$.

On sweeping the magnetic field, $E_{\mathrm{p}}$ manifests itself as an irreversible (hysteretic) contribution to the magnetization $\delta M_{\mathrm{irr}}$~\cite{goto1}. This energy must be released as heat as soon as the phase boundary is crossed, giving rise to the
irreversible contribution $\delta T_{\mathrm{irr}}$ to the MCE. Upon making
a rather simple assumption that $\delta T_{\mathrm{irr}}\propto E_{\mathrm{p}}$,
fits of $E_{\rm p}(T)$ in Fig.~\ref{analysis} reproduce the experimental data rather well for both U(Ru$_{0.96}$Rh$_{0.04}$)$_2$Si$_2$ and Yb$_{1-x}$Y$_x$InCu$_4$, yielding 
$U_0=$~1~$\pm$~0.3~K for both systems. Time-dependent magnetization measurements of 
Yb$_{1-x}$Y$_x$InCu$_4$ (with $x=$~0.1) in Fig.~\ref{metastable}b provide
rather direct evidence for metastability and creep, revealing that, after
cooling the sample part way through the transition, the magnetization
changes slowly under a constant $H$ and $T$, having an approximate
logarithmic time dependence. The large critical fields
prevent an equivalent study from being made on U(Ru$_{0.96}$Rh$_{0.04} $)$_2$Si$_2$.

% Alejandro: justification of the plot you use is needed here.

In summary, we have shown that the irreversible properties of the first
order phase transition into the field-induced phase (phase II)
of U(Ru$_{0.96}$Rh$_{0.04}$)$_2$Si$_2$ are very similar to that associated
with the valence transition of Yb$_{1-x}$Y$_x$InCu$_4$. This suggests that the broken symmetry order parameter responsible for phase II is coupled to a change in valence, or equivalently $V_{fc}$.
Given that the low $H$ Fermi liquid of URu$_2$Si$_2$ (in which field-induced phase II also occurs~\cite{kim1}) is consistent with a lowest energy $\Gamma_5$ doublet~\cite{silhanek2}, it is reasonable to expect some form of antiferroquadrupolar order. We propose that the collapse of the quasiparticle bandwidth associated with metamagnetism~\cite{silhanek1} favors field-induced local moment electric quadrupolar ordering in URu$_2$Si$_2$~\cite{ohkawa1} over the itinerant phases, which is further favored in U(Ru$_{0.96}$Rh$_{0.04}$)$_2$Si$_2$ by Rh-doping. A local moment ordering within phase II lends itself more easily to the established local spectroscopic probes such as NQR and resonant x-ray scattering~\cite{mcmorrow1}, as well as magnetoelastic techniques such as magnetostriction and thermal expansion which have yet to be performed at the high magnetic fields necessary to access phase II.

The itinerant order parameter responsible for the low $H$ HO phase in pure URu$_2$Si$_2$, in contrast, may lend itself more accessible to probes that are more suitable for studying itinerant quasiparticles, such as the dHvA effect. Such probes have already revealed the Zeeman splitting of quasiparticle with Ising spin degrees of freedom~\cite{silhanek2}, but could in principle be extended to studying the equivalent splitting of electric quadrupolar degrees of freedom by applying uniaxial strain in the appropriate X or Y direction.

%{\it acknowledgments.-} We would like to thank .................
This work was performed under the auspices of the National Science
Foundation, the Department of Energy (US) and the State of Florida.


\begin{thebibliography}{99}
\bibitem{nakashima1} Also be true for the Ising-antiferromagnetic phase observed under pressure, since the Fermi surface topology there is observed to be the same as for the HO phase; M.~Nakashima {\it et al.}, Physica B {\bf 329}, 566 (2003).
\end{thebibliography}
\end{document}